\documentclass[aps,twocolumn,prl,longbibliography]{revtex4-1}
\usepackage[english]{babel}
\usepackage{amsmath,amssymb}
\usepackage{graphicx}
\usepackage{color}

\begin{document}
\title{Efficacy of information transmission in cellular communication}
\author{Sumantra Sarkar}
\email{sumantra@lanl.gov}
\affiliation{Center for Nonlinear Studies, Theoretical Division, Los Alamos National Laboratory, Los Alamos, NM 87545, U.S.A.}

\author{Sandeep Choubey}
\email{sandeep@pks.mpg.de}
\affiliation{Max Planck Institute for the Physics of Complex Systems, Nothnitzer Strasse 38, 01187 Dresden, Germany}

\begin{abstract}
Inter and intra-cellular signaling are essential for individual cells to execute various physiological tasks and accurately respond to changes in their environment. Signaling is carried out via diffusible molecules, the transport of which is often aided by active processes that provide directional advection. How diffusion and advection together impact the accuracy of information transmission during cell signaling remains less studied. To this end, we study a one-dimensional model of cell signaling and compute the mutual information (MI) as a measure of information transmission. We find that the efficacy of the information transmission improves with advection only when  the system parameters result in Peclet number greater than one. Intriguingly, MI exhibits nontrivial scaling with the Peclet number, characterized by three distinct regimes. We demonstrate that the observed dependence of MI on the transport properties of signaling molecules has important consequences on cellular communication. 
\end{abstract}

\maketitle

\begin{figure}
\centering 
\includegraphics[width=0.5\textwidth]{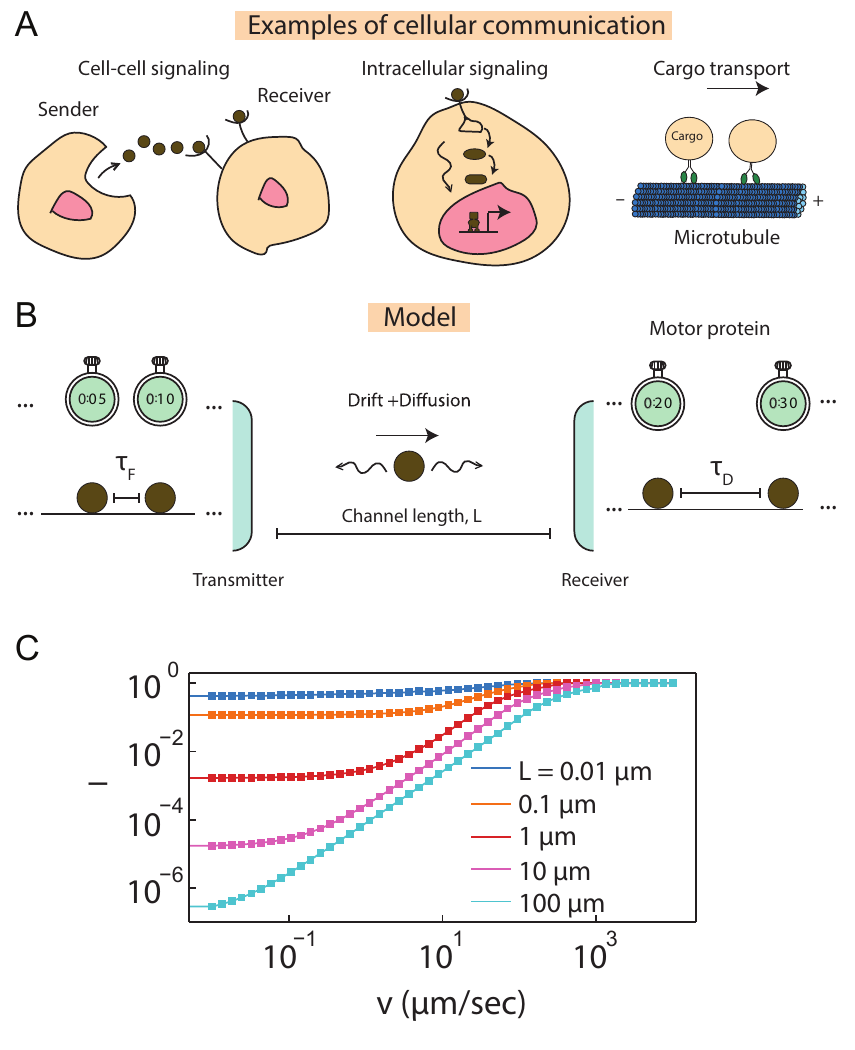}
\caption{\textbf{Model}: (A) In cellular communication channels, information is transmitted through the physical transport of signaling molecules from a transmitter to a receiver. Examples of cellular communication systems are shown. (B) We model such cellular communication systems as a one-dimensional channel with a perfectly reflecting transmitter at $x = 0$ and a perfectly absorbing receiver at $x = L$. Signaling molecules are released from the transmitter at different time points; the times between two firing events is denoted as $\tau_F$, which are drawn from a Gaussian distribution. Once emitted, these molecules traverse the channel. The transport of the signaling molecule can happen through either pure diffusion (diffusion constant $D$) or through both diffusion and advection at a constant velocity $v$ in the $+x$ direction. These molecules are absorbed at different times at the receiver. The times between two consecutive detection events is denoted by $\tau_D$. We measure the mutual information $I(\tau_F;\tau_D)$ between the distributions of the firing time interval $P(\tau_F)$ and the detection time interval $P(\tau_D)$. (C) $I$ as a function of $v$ for $D = 1 \mu m^2/s$ is shown in the bottom panel.}
\label{fig:model}
\end{figure}

The ability of individual cells to communicate, and correctly respond to any alteration in their environment forms the basis of development, immunity and tissue repair. During cellular communication an individual cell receives signals from nearby cells or surrounding environment. Upon receiving the external signal the cell transmits the decoded information about the extracellular environment to downstream effectors, which enables the cell to regulate its physiological state in response to changing environments. Such cellular communication processes are carried out by quintessential diffusible signaling molecules~\cite{phillips2012physical}. The motion of the signaling molecules is often aided by active processes that provide directional advection. Examples of such cellular communication systems include pheromone diffusion, quorum sensing, ion channel (e.g. calcium) diffusion, molecular motors carrying cargo along microtubules, etc (Fig.~\ref{fig:model}A)~\cite{phillips2012physical,lim2014cell,sarkar2020presence,ngo2020anionic}. The efficacy of cellular communication has been studied extensively using ideas from information theory~\cite{bialek2008cooperativity, de_ronde_effect_2010, tkacik_information_2016, cheong_information_2011, santos_growth_2007, selimkhanov_accurate_2014, tostevin_mutual_2009,becker2015optimal,erez2020cell}. However, most of these studies focused on biochemical networks pertaining to the various singaling pathways~\cite{hao_tunable_2013,hoffmann_ikappab-nf-kappab_2002,santos_growth_2007,purvis_encoding_2013,thurley2018modeling}. How transport properties of signaling molecules impact the accuracy of temporal information transmission remains elusive \cite{pierobon_capacity_2013,kadloor2012molecular,dieterle_dynamics_2019}. The goal of this letter is to develop a framework that allows us to study the efficacy of temporal information transmission in cellular communication.

\paragraph*{Model} We consider a simple model of cellular communication: signaling molecule moving in a one-dimensional channel of length $L$. The transmitter of the signal is located at $x = 0$, and the receiver is located at $x = L$ (Fig.~\ref{fig:model}B). In addition, we assume that the transmitter is a perfect reflector and the receiver is a perfect absorber. Hence we impose reflecting boundary condition at $x=0$ and absorbing boundary condition at $x = L$. Signaling molecules are fired at different time points from the transmitter. The times between two consecutive firing events is denoted as $t_F$. Once emitted from the transmitter, each of these molecules moves using either pure diffusion or a combination of advection and diffusion until it gets absorbed by the receiver. The times between consecutive detection events is given by $t_D$. We assume that the receiver gathers information about $\tau_F$ from $\tau_D$. Therefore, the information processing by the communication channel is intertwined with the underlying transport process, which is characterized by the free diffusion of the signaling molecules with a diffusion coefficient $D$ and a drift velocity $v$ in the positive $x$-direction.\\
To characterize the impact of transport on the efficacy of information transmission, we computed the mutual information (a well-established metric to characterize information transmission \cite{tkacik_information_2016}) between the distribution of the firing time interval $P(\tau_F)$, where $ \tau_F = t_F^{(i+1)} - t_F^{(i)}$ and the distribution of the detection time interval $P(\tau_D)$, where $ \tau_D = t_D^{(j+1)} - t_D^{(j)}$. Here $ i \mbox{ and } j $ denote the $ i^{th}$ firing event and $ j^{th}$ detection event: 
\begin{align}
I(\tau_F;\tau_D) &= \sum_{\tau_F} \sum_{\tau_D} P(\tau_F,\tau_D) \log \frac{P(\tau_F,\tau_D)}{P(\tau_F)P(\tau_D)}\\
&=  \sum_{\tau_F} \sum_{\tau_D} P(\tau_F) P(\tau_D\vert\tau_F) \log \frac{P(\tau_D\vert\tau_F)}{P(\tau_D)}\label{eqn:MI}.
\end{align}
Furthermore, without any loss of generality, we define the normalized mutual information, $I = I(\tau_F;\tau_D)/H(\tau_F)$, where $H(\tau_F)$ is the entropy of the firing time distribution. Also, without loss of generality, we assume that $P(\tau_F)$ is a gaussian. 

In general, the $i^{th}$ firing event may not lead to the $ i^{th}$ detection event, as the $i^{th}$ signaling molecule may take a more circuitous path than $ (i+1)^{th}$ signaling molecule, ending up reaching the receiver later. However, in 1D, the ordering of firing events coincide with the ordering of detection events due to excluded volume interaction of molecules in the 1D channel. We assume that the firing events are separated enough so that the influence of the excluded volume interaction on the first passage time distribution is negligible, which is further confirmed by simulations (Fig. S1). This realization simplifies the computation of mutual information significantly. Formally, we can write the detection times of two molecules 1 and 2 as:
\begin{eqnarray}
t_D^{(1)} = t_F^{(1)} + t_T^{(1)} \label{eqn:tD1}\\
t_D^{(2)} = t_F^{(2)} + t_T^{(2)}\label{eqn:tD2},
\end{eqnarray}
where $ t_F^{(2)} > t_F^{(1)}$, and $ t_T^{(1,2)}$ are the transport times of the molecules. Because the receiver in our problem is a perfect absorber, the distribution of the transport times is given by the first passage times of signaling molecules from the transmitter to the receiver, which we denote as $ F(t)$. Combining Eq.\ref{eqn:tD1} and \ref{eqn:tD2}, we find that 
\begin{eqnarray}
\tau_D = t_D^{(2)} - t_D^{(1)} = \tau_F + t_T^{(2)} - t_T^{(1)}.
\end{eqnarray}
Therefore, the distribution of the detection time interval is given by:
\begin{align}
P(\tau_D) &= \int_{-\infty}^{\infty} d\tau_F \int_{0}^{\infty} dt P(\tau_F)F(t)F(\tau_F + t - \tau_D)  \label{eqn:tD}\\
P(\tau_D|\tau_F) &= \int_{0}^{\infty}dt F(t)F(\tau_F + t - \tau_D)  \label{eqn:tD_given_tF},
\end{align}

Using $P(\tau_D)$, $P(\tau_D|\tau_F)$, and $P(\tau_F)$, we can compute the mutual information $I$. However, to calculate $P(\tau_D)$ and $P(\tau_D|\tau_F)$, we first need to find the first passage time distribution $F(t)$. In 1D, it can be found by solving the 1D diffusion equation, subjected to the appropriate boundary conditions, which yields an infinite series~\cite{redner2001guide}. Therefore, for the ease of exposition, we use an approximate functional form given by Eq.~\ref{eqn:FPTD}. The functional form of $F(t)$ used here works well for all values of the drift velocity $v$ (see SI). 
\begin{align}
F(t) &= \frac{CL}{\sqrt{4\pi Dt^3}}\exp\left[-\frac{(vt - L)^2}{4Dt}\right]\times \exp \left[-\frac{D^2t^2}{2L^4}\right]\label{eqn:FPTD},
\end{align} where $C$ is a normalization constant. 

\paragraph*{Mutual information as a function of the drift velocity} In Fig.~\ref{fig:model}C, we show $I$ computed using above equations as a function of the magnitude of the drift velocity $v$. We find that, for all channel lengths, $I$ increases as $v$ is increased. However, we find that the increase in $I$ is nonmonotonic with the drift velocity $v$. In particular, the drift velocity has to cross a threshold value, only after which $I$ changes appreciably with $v$. This result suggests that $I$ has a complex dependence on the underlying transport process and we need to carefully understand how the transport of the signaling molecules through the interplay of diffusion and advection shapes the accuracy of the transmitted information. 

\begin{figure}
\centering 
\includegraphics[width=0.5\textwidth]{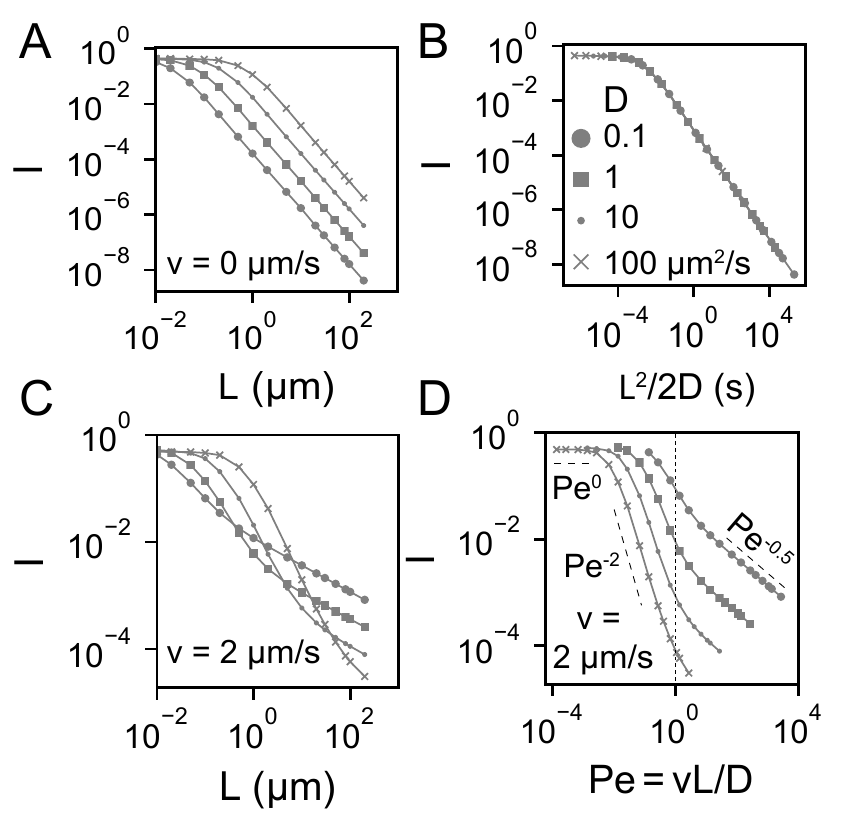}
\caption{(A) Mutual information, $I$ as a function of channel length $L$ for different diffusion coefficients (values shown in the legend of B). (B) $I$ is inversely proportional to $T = L^2/2D$ beyond a length scale $l_0$ (see text). (C) $I$ vs $L$ for different $D$ values for $v = 2 \mu m/s$ shows a third transition beyond a length scale $l_1$ (see text), where $I$ scales as $L^{-0.5}$, in addition to the transitions shown in B. (D) $I$ vs Peclet number $Pe = vL/D$ for different $D$ values at $v = 2 \mu m/s$ shows that the third transition happens when $Pe = 1$ (dashed vertical line).}
\label{fig:mi}
\end{figure}

\paragraph*{Mutual information in pure diffusive channels} In the absence of any driving, i.e. when $v = 0 \mu m/s$, transport happens purely through diffusion. For this setting, $I$ depends on the channel length $L$ and the diffusion coefficient $D$ (Fig.~\ref{fig:mi}A). In general, we find that for a given value of $D$, $I$ does not show any variation with the channel length $L$ up to a $D$-dependent length scale, $l_0$, beyond which $I$ decays in a universal fashion across different diffusion coefficients. In fact, we find that in this regime, $I$ is inversely proportional to $T = L^2/2D$ (Fig.~\ref{fig:mi}B), the typical time (the sojourn time) between two consecutive detection events for 1D channels of length $L$. This observation suggests that the cellular communication process undergoes a transition at the threshold $l_0$: below $l_0$ variation of $I$ depends only on $P(\tau_F)$ and the transport process does not influence its variation, whereas, the variation of $I$ above $l_0$ is strongly dependent on the transport process and $P(\tau_F)$ plays a secondary role. To estimate $l_0$, we note that transport becomes important when the sojourn time is comparable to the average time interval between consecutive firing events, $\langle\tau_F\rangle$, such that $l_0 \sim \sqrt{2D\langle\tau_F\rangle}$.

\begin{figure}
\centering 
\includegraphics[width=0.5\textwidth]{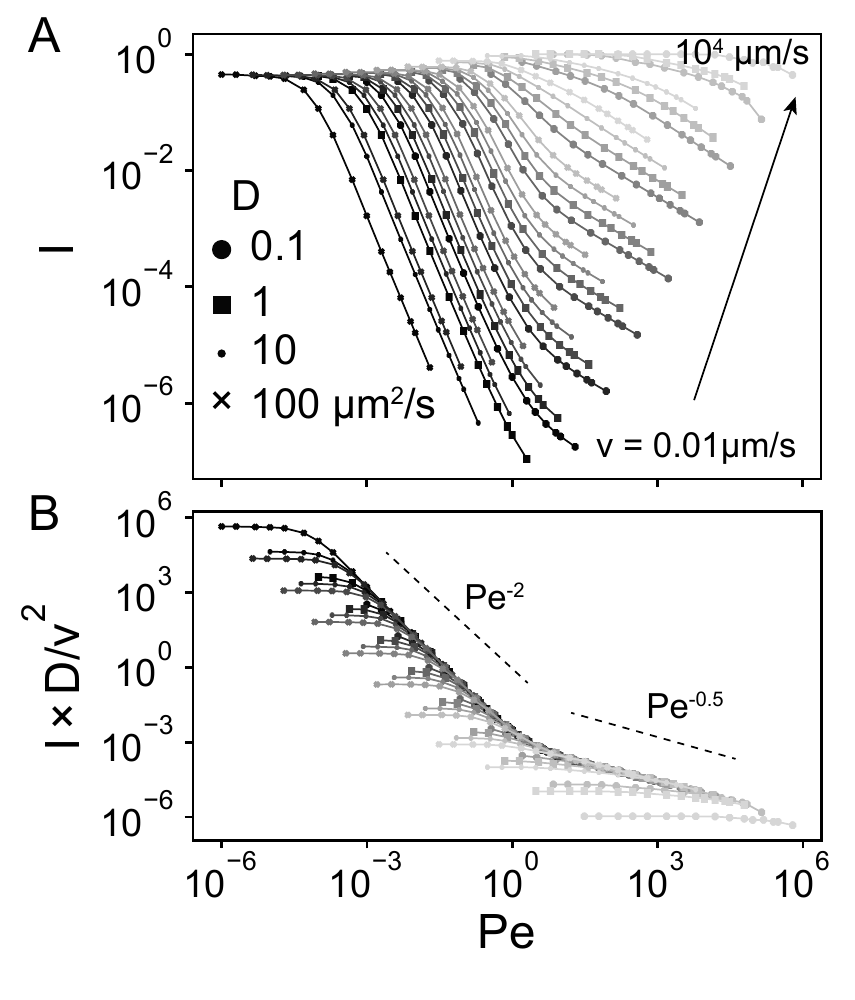}
\caption{(A) $I$ vs $Pe$ for different values of the drift velocities and diffusion coefficients (legends). (B) $I\times D/v^2$ vs $Pe$ shows scaling collapse on to a single master curve that scales as $Pe^{-2}$ for $Pe < 1$ and as $Pe^{-0.5}$ for $Pe > 1$. }
\label{fig:scaling}
\end{figure}

\paragraph*{Mutual information in active channels} To explore the impact of nonzero driving we computed the variation of $I$ with $L$ when $v > 0$. We observe that the transition at $l_0$ persists even in the presence of nonzero $v$ (Fig.~\ref{fig:mi}C). Furthermore, we find that the introduction of drift leads to another transition at another $D$-dependent characteristic length scale $l_1$. For $L > l_1$, $I$ decays as $L^{-0.5}$, which is slower than the $L^{-2}$ decay observed when $ l_0 < L < l_1 $. Because the $L^{-2}$ variation stems from the variation of the diffusive sojourn times with $L$, we suspected the transition at $l_1$ originates from the introduction of the drift. Indeed, when we measured the variation of $I$ with the Peclet number, $Pe = \frac{vL}{D}$, we found that the transition to the $L^{-0.5}$ (equivalently $Pe^{-0.5}$) decay regime occurs precisely at $Pe = 1$, where the advection rate $vL$ is exactly equal to the diffusion coefficient (Fig.~\ref{fig:mi}D). This observation also implies that $l_1 = D/v$.  Furthermore, as $v$ is increased, there is a transition from a region where $Pe < 1$ even at the largest channel length probed ($L=200 \mu m$, e.g. length of a neuron), such that the decay of $I$ occurs only through the passive diffusion process, to a region where $Pe > 1$ even at the smallest channel length probed ($L = 0.01 \mu m$), such that the decay of $I$ is mediated by only the active advection process (Fig.~\ref{fig:scaling}A). Furthermore, when $I$ is multiplied by a timescale $D/v^2 = l_1/v$, $I(Pe)$ collapse onto a single master curve that scales as $Pe^{-2}$ below $Pe < 1$ and as $Pe^{-0.5}$ for $Pe > 1$ (Fig.~\ref{fig:scaling}B). Such a scaling is observed because beyond $l_0$, variation in MI is solely determined by the transport process (SI). In particular, we find that beyond this length scale, MI is inversely proportional to the sojourn time between two detection events, which shows exactly the same scaling behavior with Pe.

\begin{figure}
\centering 
\includegraphics[width=0.5\textwidth]{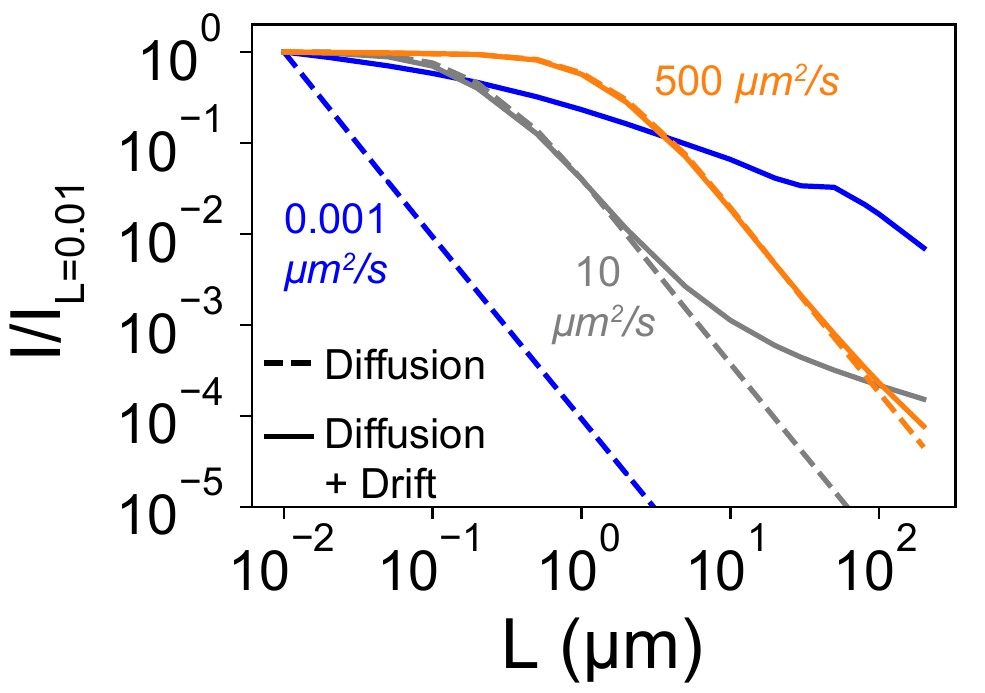}
\caption{The mutual information, $I$ as a function of channel length, $L$, for diffusive and advective (diffusion + drift, $v = 2 \mu m/s$) channels. To better demonstrate the decay of $I$ with $L$, we have normalized $I$ by the value it takes at the smallest length scale ($L = 0.01 \mu m $). If the signaling molecules are second messengers molecules (orange) or proteins (grey), for small $L$ values such that $Pe < 1$, there is negligible difference between the decay of $I$ with $L$ in the diffusive and the advective channels respectively. However, if the signaling molecules are vesicles (blue), then $I$ decays faster in the diffusive channel than in the advective channel.}
\label{fig:signaling}
\end{figure}

\paragraph*{Impact on cell signaling} Results in the previous two paragraphs suggests that the transport of the signaling molecules plays a pivotal role in determining the accuracy of the transmitted signals. In particular, we found that the decay of mutual information is characterized by two length scales $l_0$ and $l_1$. The diffusive transport process becomes important at $l_0$, whereas driven transport through advection becomes important at $l_1$. Because $l_0$ is usually smaller than $l_1$, it implies that the addition of driving may not lead to an immediate increase in the accuracy of the transmitted signals. Because driving leads to improvement in accuracy only when $Pe > 1$, immediate improvement is possible if and only if $l_1 \leq l_0$, so that $v^2 \geq D/2\langle \tau_F\rangle$ or $D \leq v^2\langle \tau_F\rangle$. To develop some intuition about these thresholds, let's consider typical values of these variables encountered in cell-signaling processes. At $nM-\mu M$ concentrations of the ligands, the typical interval between two firing events, $\langle \tau_F\rangle$, are of the order of  $10 ms$ or $0.01 s$~\cite{milo2010bionumbers}, whereas the maximum drift velocity encountered in cellular systems is of the order of $1 \mu m/s$~\cite{milo2010bionumbers}. Therefore, drift will lead to immediate improvement when $D \leq 0.01 \mu m^2/s$, the typical diffusion coefficients for large assemblies, such as vesicles~\cite{milo2010bionumbers}. Similar calculations show that if $D = 10 \mu m^2/s$, which is the typical diffusion coefficient of proteins in the cytosol~\cite{milo2010bionumbers}, $v$  has to be greater than $30 \mu m/s$, and for second messengers with $D \sim 500 \mu m^2/s$,  $v$  has to be greater than $200 \mu m/s$, both of which are unfeasible in the biological setting~\cite{milo2010bionumbers}. Indeed, when we compare the information loss during diffusive transport and driven transport with biologically relevant drift velocity, we find that the results echo the foregoing discussion (Fig.~\ref{fig:signaling}), with drift dominating diffusion for all channel lengths only when $D < 0.01 \mu m^2/s$.

\paragraph*{Discussion} In this letter, we study the effect of molecular transport on the efficacy of information transmission during cellular communication. In particular, we consider two scenarios: pure diffusion and diffusion with advection. Using the mutual information as a metric for accuracy of information transmission, we find that MI exhibits a nonlinear decrease as a function of the channel length in both scenarios. For pure diffusion, initially for small channel length, MI doesn’t show any significant drop. However, for longer channel lengths it decreases sharply. Notably, through a change in variable, all the curves for mutual information as a function of channel length collapse onto a single master curve, showing a scaling of $L^{-2}$. When advection is added to the model, MI initially remains constant for small drift; further increase in the drift velocity leads to an increase in the MI, which subsequently plateaus. This effect can be characterized better when MI is plotted against the Peclet number; MI shows three distinct regimes with three different scaling factors. We discover that drift becomes important only when the Peclet number is above one, at which point the advection is dominant over diffusion. For $Pe < 1$, drift doesn’t improve the efficacy of information transmission. Our results demonstrate nontrivial dependence of information transmission on the transport properties of signaling molecules. A renewed look into biological systems in light of these results provides interesting insights. For cell signaling through proteins and small molecules, the diffusion coefficients are of the order of $10 -100 \mu m^2/s$, which transmit signal over a length of $0.1-1 \mu m$. Therefore advection can improve the accuracy of the transmitted signals if the drift velocity is of the order of $10-100 \mu m/s$, which is much larger than that of typical molecular motors which have drift velocities of around $1 \mu m/s$~\cite{milo2010bionumbers}. Therefore, from the perspective of information transmission, for small molecule or protein driven cell signaling, the preferred mode of transport will be diffusion. Whereas, for signaling through large assemblies, such as vesicles, that have diffusion coefficients of the order of $0.001 \mu m^2/s$~\cite{milo2010bionumbers}, drift is important for any channel of length greater than 1 nm. Therefore, advection might be the preferred mode of transport for neuronal signaling, where information is transported through synaptic vesicles. Since information transmission is a fundamental function of cellular communication networks, an exciting plausibility is that evolutionary pressures would shape the cellular machinery to maximize the reliable decoding of temporal signals.

\begin{acknowledgements}
The authors thank Sidney Redner, Angel E. Garcia, Arjun Narayanan, Md. Zulfikar Ali for useful comments and critical reading of the manuscripts. SS was funded by an LDRD grant (XX01) from LANL. This paper has the following LA-UR number: LA-UR-20-28073.
\end{acknowledgements}
\bibliography{main}
\widetext
\clearpage
\twocolumngrid
\begin{center}
\textbf{\large Supplementary Information: Efficacy of information transmission in cellular communication}
\end{center}
\setcounter{equation}{0}
\setcounter{figure}{0}
\setcounter{table}{0}
\setcounter{page}{1}
\makeatletter
\renewcommand{\theequation}{S\arabic{equation}}
\renewcommand{\thefigure}{S\arabic{figure}}

\section{First passage time distribution}

\begin{figure*}
\centering 
\includegraphics[width=\textwidth]{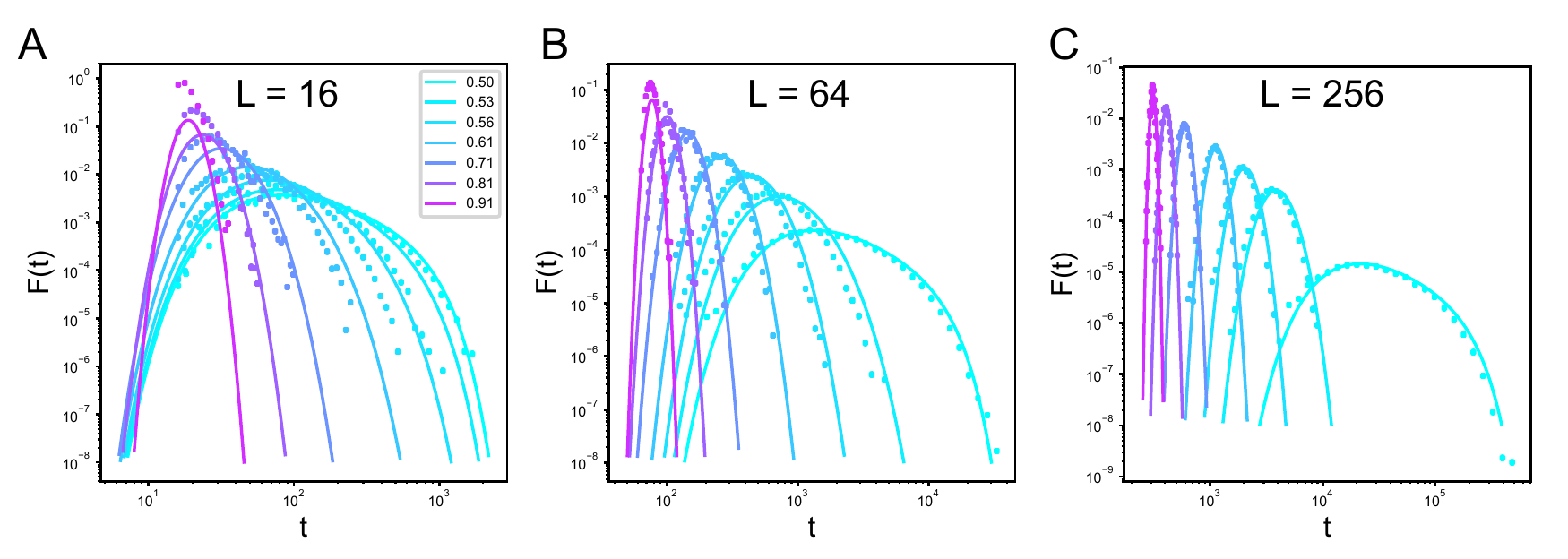}
\caption{First passage time distribution obtained from lattice model (dots) and Eq.~\ref{eqn:FPTD} (solid line) for (A) $L = 16$, (B) $L = 64$, (C) $L = 256$. }
\label{fig:SI-1}
\end{figure*}

We model a cellular communication channel as a 1D finite region of length $L$ with reflecting boundary condition at $x = 0$ and absorbing boundary condition at $x = L$. This set up is referred to as the transmission mode in ref~\cite{redner2001guide}. The first passage time distribution for this problem is given by the first passage time distribution for free diffusion in infinite space added to contribution from infinite image charges originating from the reflecting and the absorbing boundary conditions, resulting in an infinite sum of terms~\cite{redner2001guide}. 

To avoid convergence issues of the infinite series during numerical evaluation, we use an approximate formula for the first passage time distribution. To construct this formula, we note that in the presence of large enough drift in the $+x$ direction, the presence of the reflecting boundary is rarely felt by the transported molecule. Therefore, the first passage time distribution is chosen to be of the following form: 
\begin{align}
    F(t) &= F_{abs}(t)*g(t) \label{eq:FT}\\
    F_{abs}(t) &=  \frac{L}{\sqrt{4\pi Dt^3}}\exp\left[-\frac{(vt - L)^2}{4Dt}\right] , 
\end{align}
where $F_{abs}$ is the first passage distribution of a molecule transported in a semi-infinite line with absorbing boundary condition at $x = L$~\cite{kadloor2012molecular,redner2001guide}, and $g(t)$ is a function that captures the contribution of the reflecting boundary condition at $x = L$.  

To determine $g(t)$ we simulate a 1D lattice model of length $L$ with same boundary condition. In the lattice model the particle hops with a constant rate $p$ in the $+x$ direction and with a rate $q$ in the $-x$ direction. When $p=q$, the particle motion is purely diffusive and when $p$ and $q$ are unequal, the particle moves using both advection and diffusion. For when $p>q$, the particle has an effective drift velocity of $a(p - q)$, where $a$ is the length of each lattice site. The corresponding diffusion coefficient is $D=a^2(p+q)/2$~\cite{phillips2012physical}. From the simulations, the  measured first passage time distribution has a gaussian tail (Fig.~\ref{fig:SI-1}), and we find that the following ansatz approximate the true distribution well:  

\begin{align}
F(t) &= \frac{CL}{\sqrt{4\pi Dt^3}}\exp\left[-\frac{(vt - L)^2}{4Dt}\right]\times \exp \left[-\frac{D^2t^2}{2L^4}\right]\label{eqn:FPTD},
\end{align} where $C$ is a normalization constant. We use this approximate formula because the first two moments of the distribution are nearly identical to the true distribution~\cite{redner2001guide}. In fact, this ansatz was motivated by the analytical expressions of the first two moments derived in~\cite{redner2001guide}. This result implies that $g(t) = \exp \left[-\frac{D^2t^2}{2L^4}\right]$. To measure the first passage distribution using the lattice model we generated 10000 trajectories at each $p$ values ($q = 1 - p$). As Fig.~\ref{fig:SI-1} shows, this approximation works quite well (Peclet number varies from 0 to $4\times10^2$).  

\section{Mean sojourn time}

\begin{figure}
\centering 
\includegraphics[width=0.5\textwidth]{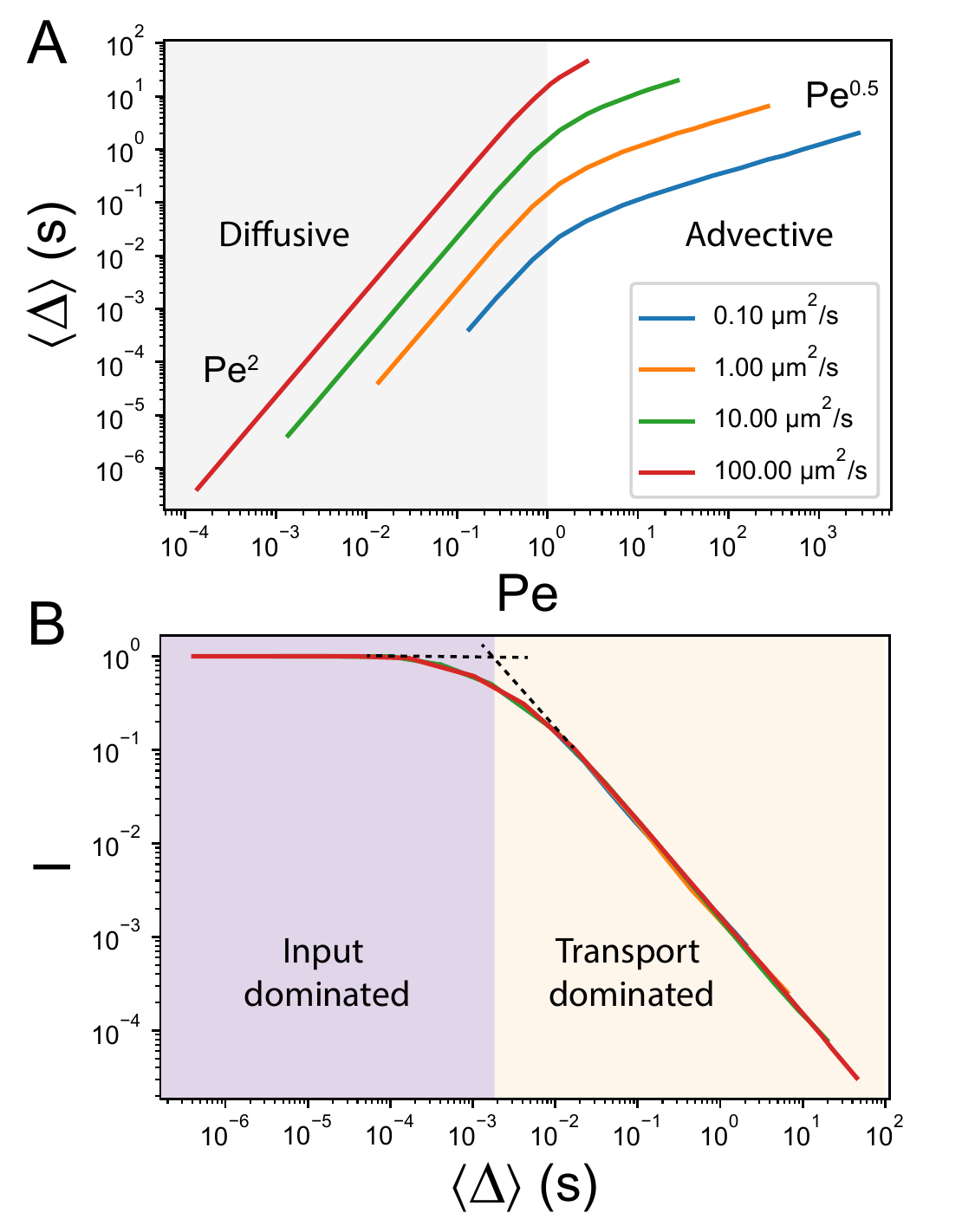}
\caption{(A) $\langle\Delta\rangle$ vs $Pe$ for different values of diffusion coefficients (legends) at drift velocity of $v = 2\mu m^2/s$. (B) $I$ vs $\langle \Delta \rangle$ shows that when  $\langle \Delta \rangle >> \langle \tau_F \rangle = 10^{-2}$ , $I$ is inveresely proportional to $\langle \Delta\rangle$. }
\label{fig:SI-2}
\end{figure}

As shown in the main text the time interval between two consecutive detection event is given by: 

\begin{align}
\tau_D &= t_D^{(2)} - t_D^{(1)} = \tau_F + t_T^{(2)} - t_T^{(1)} \\
\tau_D &= \tau_F + \Delta \label{eq:tD},  
\end{align}
where $\Delta$ is the difference between the transport time of two consecutive molecules detected at the receiver. As Fig.~\ref{fig:SI-2}A shows the average transport time difference $\langle \Delta \rangle$ scales as $Pe^2$ when $Pe < 1$ and as $Pe^{0.5}$ for $Pe > 1$. This scaling is exactly inverse of how $I$ varies with $Pe$ in Fig. 3, which implies that $I$ is inversely proportional to $\langle \Delta \rangle$. Indeed, As Eq.~\ref{eq:tD} suggests, when $\tau_F >> \Delta$, $\tau_D$ is determined by the firing time interval distribution $\tau_F$ (the input distribution). On the other hand, when $\Delta >> \tau_F $, $\tau_D$ is determined by the transport process (Fig.~\ref{fig:SI-2}B).

\end{document}